\documentclass{PoS}

\usepackage{amsmath,amssymb}
\usepackage{graphicx}

\newcommand{\p}{\partial}
\newcommand{\pslash}{p\kern-1ex /}
\newcommand{\lslash}{l\kern-1ex /}
\newcommand{\kslash}{k\kern-1ex /}
\newcommand{\dslash}{\p\kern-1.2ex /}
\newcommand{\Dslash}{{\cal D}\kern-1.5ex /}
\newcommand{\Aslash}{A\kern-1.2ex /}

\newcommand{\tr}{{\rm tr}}
\newcommand{\Tr}{{\rm Tr}}

\newcommand{\bea}{\begin{eqnarray}}
\newcommand{\eea}{\end{eqnarray}}

\newcommand{\nn}{\nonumber\\}

\newcommand{\BAN}{\begin{eqnarray*}}
\newcommand{\EAN}{\end{eqnarray*}}
\title{Chiral symmetry and axial U(1) symmetry in finite temperature QCD with domain-wall fermion}

\ShortTitle{Chiral symmetry and axial U(1) symmetry in finite temperature QCD}

\author{\speaker{Ting-Wai Chiu}$^{1,2,3}$, 
        Wen-Ping Chen$^{1}$, Yu-Chih Chen$^{1}$, Han-Yi Chou$^{1}$, Tung-Han Hsieh$^4$ (TWQCD Collaboration)\\
$^1$ Physics Department, National Taiwan University, Taipei 10617, Taiwan\\
$^2$ Center for Quantum Science and Engineering, National Taiwan University,\\
     \hspace{2.5mm}Taipei 10617, Taiwan\\
$^3$ Center for Theoretical Sciences, National Taiwan University,
     Taipei 10617, Taiwan\\
$^4$ Research Center for Applied Sciences, Academia Sinica, Taipei 115, Taiwan}

\abstract{

We study the restoration of the spontaneously broken chiral symmetry and the anomalously broken 
axial U(1) symmetry in finite temperature QCD at zero chemical potential. 
We use 2 flavors lattice QCD with optimal domain-wall fermion on the $ 16^3 \times 6 $ lattice,  
with the extent $ N_s = 16 $ in the fifth dimension, in the temperature range $ T = 130-230 $ MeV.
To examine the restoration of the chiral symmetry and the axial $ U(1) $ symmetry, 
we use diluted $ Z_2 $ noises to calculate the chiral condensate, and the chiral susceptibilities 
in the scalar and pseudoscalar meson channels, for flavor singlet and non-singlet respectively. 
From the degeneracy of the chiral susceptibilities around $ T_c $, it suggests that the axial 
$ U(1) $ symmetry is restored in the chirally symmetric phase.  
Moreover, we examine the spectral density $ \rho(\lambda_c) $ of the 4D effective Dirac operator with exact chiral symmetry,    
which is obtained by computing zero modes plus (180+180) conjugate pairs of low-lying modes for each gauge configuration. 
The suppression of low modes in the spectral density provides a consistency check of the restoration 
of axial $ U(1) $ symmetry in the chirally symmetric phase.   
}

\FullConference{31st International Symposium on Lattice Field Theory LATTICE 2013\\
                 July 29 - August 3, 2013\\
                 Mainz, Germany}

\begin{document}

\section{Introduction}

In QCD, the classical action of $ N_f $ massless quarks has the symmetry 
$ SU(N_f)_L \times SU(N_f)_R \times U(1)_V \times U(1)_A $. 
In the quantum theory, at zero temperature, the chiral symmetry $ SU(N_f)_L \times SU(N_f)_R $
is broken spontaneously to $ SU(N_f)_V $ by the vacuum of QCD, and the $ U(1)_A $ symmetry is broken 
by the axial anomaly. It is expected that at high temperature, both chiral symmetry and $ U(1)_A $ symmetry
are restored. The question is, at what temperature $ T_c $ the chiral symmetry is restored, 
and whether $ U(1)_A $ symmetry is also restored at $ T_1 \simeq T_c $. 

Lattice QCD with exact chiral symmetry \cite{Kaplan:1992bt,Neuberger:1997fp}
is an ideal theoretical framework to study the nonperturbative physics 
from the first principles of QCD. Thus it is in a good position to answer above 
questions. However, it is rather nontrivial to perform Monte Carlo simulation 
such that the chiral symmetry is preserved at a high precision 
and all topological sectors are sampled ergodically. 
Currently, there are three groups (HotQCD, JLQCD, TWQCD) 
performing large-scale simulations of finite-temperature QCD with domain-wall/overlap fermion.
While HotQCD and JLQCD have been using IBM Blue Gene supercomputers, 
TWQCD has been using a GPU cluster (currently consisting 
of $ 320 $ Nvidia GPUs, with sustained $ 100 $ Tflop/s). 

The HotQCD Collaboration 
has been using the conventional domain-wall fermion with the Shamir kernel, 
which suffers from large chiral symmetry breaking 
(i.e., large residual mass), especially in the finite temperature QCD \cite{Bazavov:2012qja}. 
On the other hand, the JLQCD Collaboration and the BMW Collaboration 
have used the overlap fermion in a fixed topology, which attains very good chiral symmetry,  
but in the expense of sampling all topological sectors ergodically \cite{Cossu:2013uua, Borsanyi:2012xf}.
To overcome the deficiencies of above two approaches, TWQCD collaboration has been using 
the optimal domain-wall fermion (ODWF) \cite{Chiu:2002ir,Chiu:2009wh}
to preserve the chiral symmetry, which not only attains a good chiral symmetry
with a modest extension (e.g., $N_s=16$) in the fifth dimension, 
but also samples all topological sectors ergodically.
        
Mathematically, ODWF is a theoretical framework to preserve 
the chiral symmetry optimally with a set of analytical weights, 
$ \{ \omega_s, s = 1, \cdots, N_s \} $, 
one for each layer in the fifth dimension \cite{Chiu:2002ir}. 
Thus the artifacts due to the chiral
symmetry breaking with finite $ N_s $ can be reduced to the minimum, 
especially in the chiral regime.
In general, the 4-dimensional effective Dirac operator of massless ODWF 
can be written as \cite{Chen:2012jya}
\bea
\label{eq:odwf_4d}
\begin{aligned}
D &= [1+ \gamma_5 S_{opt}(H) ]/(2r), 
\hspace{5mm}
S_{opt}(H) = \frac{1-\prod_{s=1}^{N_s} T_s}{1 + \prod_{s=1}^{N_s} T_s}, \\
T_s &= \frac{1-\omega_s H}{1+\omega_s H},   
\hspace{5mm}
H = c H_w ( 1 + d \gamma_5 H_w)^{-1}, 
\hspace{5mm}
r = [2 m_0 (1-d m_0) ]^{-1},  
\end{aligned}
\eea
where $ c $ and $ d $ are constants, and 
$ H_w = \gamma_5 D_w (-m_0) $, with $ D_w(-m_0) $ the usual Wilson-Dirac operator
plus a negative parameter $ -m_0 ( 0 < m_0 < 2 ) $. 
Here $ S_{opt}(H) = H R_Z(H) $, where $ R_Z(H)$ is the Zolotarev optimal 
rational approximation of $ (H^2)^{-1/2} $ \cite{Chiu:2002eh}. 

Recently we have demonstrated that it is feasible to perform a large-scale dynamical QCD 
simulation with ODWF, which not only preserves the chiral symmetry to a good precision, 
but also samples all topological sectors ergodically.   
To recap, we perform HMC simulations of 2 flavors QCD on a $ 16^3 \times 32 $ lattice, 
with ODWF at $ N_s = 16 $ and plaquette gauge action at $ \beta = 5.95 $. 
Our results of the topological susceptibility \cite{Chiu:2011dz}   
and the mass and decay constant of the pseudoscalar meson \cite{Chiu:2011bm}
agree with the sea-quark mass dependence predicted by the NLO ChPT. 
This asserts that the nonperturbative chiral dynamics 
of the sea-quarks are well under control in our HMC simulations. 
Recently we have extended our simulations to larger lattices ($ 20^3 \times 40 $, $ 24^3 \times 48 $), 
with plaquette gauge action at $ \beta = 5.95 $ and $ \beta = 6.00 $ respectively.  
In this paper, we study the restoration of the spontaneously broken chiral symmetry 
and the anomalously broken axial U(1) symmetry in finite temperature QCD at zero chemical potential. 

\section{Gauge ensembles}

We perform simulations of two flavors QCD on the $16^3 \times 6$ lattice, 
with the plaquette gauge action, for 7 values of $ \beta = 6/g^2 $,  
ranging from $ \beta = 5.85 $ to $ \beta = 5.90 $.
For the quark part, we use ODWF with $ c = 1 $, $ d = 0 $ (i.e., $ H = H_w $), 
$ N_s = 16 $, and $ \lambda_{min}/\lambda_{max} = 0.01/6.2 $.
At each $ \beta $, simulations are performed for 3-4 different sea-quark masses 
$ m_q a = $ 0.01, 0.015, 0.02, and 0.03, such that extrapolation to the chiral
limit can be carried out.   
Moverover, in order to fix the scale, we also perform 
zero temperature simulations on the $ 16^3 \times 32 $ lattice, 
for $ \beta = 5.95, 5.90 $ and $ 5.88 $. 
For each ($\beta$, $m_q $) ensemble on the $ 16^3 \times 6 $ lattice,  
we generate around 4000-6000 trajectories with a single GPU or a set of GPUs. 
After discarding the initial 300 trajectories for thermalization, we sample 
one configuration every 10-20 trajectories, then we have 300-600 configurations 
for each ensemble. 

To determine the lattice scale, we compute the Wilson flow \cite{Luscher:2010iy} 
of each configuration of the zero temperature gauge ensembles at 
$ \beta$ = 5.88, 5.90, and 5.95, and use the BMW scheme \cite{Borsanyi:2012zs}
\BAN
\label{eq:w0}
\left. t \frac{d}{dt} \{ t^2 \langle E(t) \rangle \} \right|_{t=w_0^2} = 0.3, 
\EAN 
to obtain the value of $ w_0/a $ for each ensemble. 
Using the inverse lattice spacing $ a^{-1} $ we have determined at $ \beta = 5.95 $ 
(by heavy quark potential with Sommer parameter $ r_0 = 0.49 $~fm) \cite{Chiu:2011bm}, 
we obtain $ w_0 $, and also the values of $ a^{-1} $ for $ \beta = 5.90 $ and $ \beta = 5.88 $.
Then the values of $ a^{-1} $ at other $ \beta $'s are obtained by RG extrapolation.    
 
To measure the chiral symmetry breaking due to finite $N_s$, we compute the residual mass 
according to the formula \cite{Chen:2012jya}
\bea
M_{res} = \frac{ \Tr(D_c + m_q)^{-1} }{ \Tr[\gamma_5 (D_c + m_q) \gamma_5 (D_c+m_q)]^{-1} } - m_q,
\label{eq:Mres}
\eea
where 
$ (D_c + m_q)^{-1} $ denotes the valence quark propagator with $ m_q $ equal to the sea-quark mass, 
Tr denotes the trace running over the site, color, and Dirac indices.
For an ensemble of gauge configurations, one has two different ways to measure the residual mass. 
One way is to compute (\ref{eq:Mres}) for each gauge configuration of the ensemble and then obtain the average, 
which is denoted as $ \left< M_{res} \right>_1 $. 
Another way is to compute the ensemble average of the numerator and the denominator of (\ref{eq:Mres}) respectively, 
and then obtain their ratio, which is denoted as $ \left<  M_{res} \right>_2 $. Using 240 $ Z_2 $ noise vectors with dilution 
in the color and Dirac spaces ($ 20 \times 3 \times 4 = 240 $) for each configuration, 
we compute the numerator and the denominator of (\ref{eq:Mres}), and obtain the residual masses
$ \left< M_{res} \right>_1 $ and $ \left< M_{res} \right>_2 $, as listed in Table \ref{tab:Mres}. 
We see that these two different measures of the residual mass are in good agreement with each other.

\begin{table}[htbp]
\caption{Residual masses of the gauge ensembles, where the first number in each entry is $ \left< M_{res} \right>_1 $, 
         and the second number $ \left< M_{res} \right>_2 $.
         The number of configurations in each gauge ensemble varies from 300 to 600.} 
\begin{center}
\begin{tabular}{|c||c|c|c|}
\hline
   $ \beta $  & $m_q a = 0.01$ & $ m_q a = 0.02 $ & $ m_q a = 0.03 $ \\ 
\hline
\hline
      5.850 &  0.00233(12), 0.00240(13) &  0.00208(27), 0.00212(27) &  0.00151(11), 0.00151(11)   \\
      5.860 &  0.00234(12), 0.00241(12) &  0.00178(9),  0.00180(9)  &  0.00153(29), 0.00155(29)   \\
      5.870 &  0.00141(12), 0.00145(13) &  0.00155(9), 0.00157(9)   &  0.00136(6), 0.00138(6)     \\
      5.875 &  0.00062(4), 0.00064(4) &  0.00132(8), 0.00136(9) &  0.00105(5), 0.00107(5)  \\
      5.880 &  0.00094(5), 0.00097(5)    &  0.00067(5), 0.00069(5)   &  0.00074(4), 0.00075(4) \\	
      5.890 &  0.00046(5), 0.00048(5) &  0.00035(5), 0.00035(5) &  0.00041(6), 0.00042(6)  \\
      5.900 &  0.00018(2), 0.00018(2) & 0.00016(4), 0.00017(4) &  0.00022(4), 0.00022(4)  \\
\hline
\end{tabular}
\end{center}
\label{tab:Mres}
\end{table}

\section{Chiral susceptibilities}

We use the two-point functions of scalars and pseudoscalars for probing the restoration of 
$ SU(2)_L \times SU(2)_R $ symmetry, as well as the restoration of $ U(1)_A $ symmetry. 
For two flavors QCD ($ m_u = m_d = m_q $), 
in the scalar channel, we have the flavor singlet, $ \sigma = \frac{1}{\sqrt{2}} (\bar u u + \bar d d ) $,   
and flavor non-singlet, $ \delta = \bar u d, \bar d u, \frac{1}{\sqrt{2}} (\bar u u - \bar d d ) $.   
Their two-point functions are
\bea
\label{eq:Cdelta}
C_{\delta}(x) &=&  \left< (\bar u d )^\dagger (x) \bar u d (0) \right> 
              = - \left< \tr \left[ (D_c + m_q)^{-1}_{0,x} (D_c + m_q)^{-1}_{x,0} \right] \right>, \  \\ 
\label{eq:Csigma}
C_{\sigma}(x) &=&  \left< \sigma^\dagger (x) \sigma(0) \right>      
              = - \left< \tr \left[ (D_c + m_q)^{-1}_{0,x} (D_c + m_q)^{-1}_{x,0} \right] \right> \nn
              & & + 2 \left< \tr (D_c + m_q)^{-1}_{x,x} \cdot \tr (D_c + m_q)^{-1}_{0,0} \right>
                  - 2 \left< \tr (D_c + m_q)^{-1}_{x,x} \right> \left< \tr (D_c + m_q)^{-1}_{0,0} \right>, 
\eea
where the last term is added explicitly to subtract the vacuum contribution. 
The corresponding chiral susceptibilities are
\bea
\chi_\delta &=& \sum_x C_\delta(x) 
            = - \sum_x \left< \tr \left[ (D_c + m_q)^{-1}_{0,x} (D_c + m_q)^{-1}_{x,0} \right] \right> 
            = - \frac{1}{L_x^3 L_t } \left< \Tr (D_c + m_q)^{-2} \right>,  \\
\chi_\sigma &=& \sum_x C_\sigma(x) = \chi_\delta  
                + \frac{2}{L_x^3 L_t } \left\{  \left< [\Tr (D_c + m_q)^{-1}]^{2} \right>  
                                              - \left< \Tr (D_c + m_q)^{-1} \right>^2 \right\}, \\ 
\chi_{disc} &\equiv&
                 \frac{1}{L_x^3 L_t } \left\{  \left< [\Tr (D_c + m_q)^{-1}]^{2} \right>  
                                              - \left< \Tr (D_c + m_q)^{-1} \right>^2 \right\}
               = (\chi_\sigma - \chi_\delta)/2, 
\eea
where the trace Tr sums over color, Dirac, and site indices.
Similarly, in the pseudoscalar channel, we have the flavor non-singlet,  
$ \pi = \bar u \gamma_5 d, \bar d \gamma_5 u, \frac{1}{\sqrt{2}} (\bar u \gamma_5 u - \bar d \gamma_5 d ) $,   
and the flavor singlet, $ \eta = \frac{1}{\sqrt{2}} (\bar u \gamma_5 u + \bar d \gamma_5 d ) $,   
and their corresponding chiral susceptibilities, 
\bea
\label{eq:chi_pi}
\chi_\pi &=& \sum_x C_\pi(x) = \frac{1}{L_x^3 L_t } \left< \Tr [ \gamma_5 (D_c + m_q)]^{-2} \right>,  \\
\label{eq:chi_eta}
\chi_\eta &=& \sum_x C_\eta(x) = \chi_\pi  
                - \frac{2}{L_x^3 L_t } \left\{  \left< [\Tr \gamma_5 (D_c + m_q)^{-1}]^{2} \right>  
                                               -\left< \Tr \gamma_5 (D_c + m_q)^{-1} \right>^2 \right\}, \\ 
\chi_{5,disc} &\equiv&
                 \frac{1}{L_x^3 L_t } \left\{  \left< [\Tr \gamma_5 (D_c + m_q)^{-1}]^{2} \right>  
                                              - \left< \Tr \gamma_5 (D_c + m_q)^{-1} \right>^2 \right\}
               = (\chi_\pi - \chi_\eta)/2. 
\eea


Since the scalar and pseudoscalar correlation functions are related 
by $ SU_L(2) \times SU_R(2) $ flavor transformation, 
the restoration of chiral symmetry implies that 
\bea
\chi_\pi = \chi_\sigma, \quad  \chi_\delta = \chi_\eta,  
\eea
which in turn gives $ \chi_{disc} = \chi_{5,disc} $ for $ T > T_c $ and $ m_q \to 0 $.

Since $ U_A(1) $ transformation does not change the flavor quantum numbers, the restoration of $ U_A(1) $ symmetry implies 
\bea
\chi_\pi = \chi_\delta, \quad \chi_\sigma = \chi_\eta. 
\eea
If $ U(1)_A $ is restored at $ T_1 = T_c $, then $ \chi_\pi = \chi_\delta = \chi_\sigma = \chi_\eta $ for $ T > T_c $, 
and 
\bea
(\chi_\pi - \chi_\eta)
\xrightarrow{m_q \to 0} 
\left\{
\begin{matrix}
0, \quad T > T_c,     \\
\sim \frac{1}{m_q}, \quad T < T_c. \\
\end{matrix} 
\right.
\eea
If $ U_A(1) $ symmetry is broken above $ T_c $, then there exists a window $ T_c < T <  T_1 $ in which 
$ \chi_\pi = \chi_\sigma $ and $ \chi_\delta = \chi_\eta $, but $ \chi_\pi \neq \chi_\delta $ and $ \chi_\sigma \neq \chi_\eta $. 
If the chiral symmetry restoration (phase transition) belongs to the $ O(4) $ universality class, 
then we expect 
\bea
(\chi_\pi - \chi_\eta)
\sim (T-T_c)^{-\gamma},  \ \gamma=1.453, 
\eea
for $ T_c < T < T_1 $ and $ m_q \to 0 $. 

\section{Preliminary results}

\begin{figure}[!htb]
\begin{center}
\begin{tabular}{@{}cccc@{}}
\includegraphics*[height=4cm,width=4.8cm,clip=true]{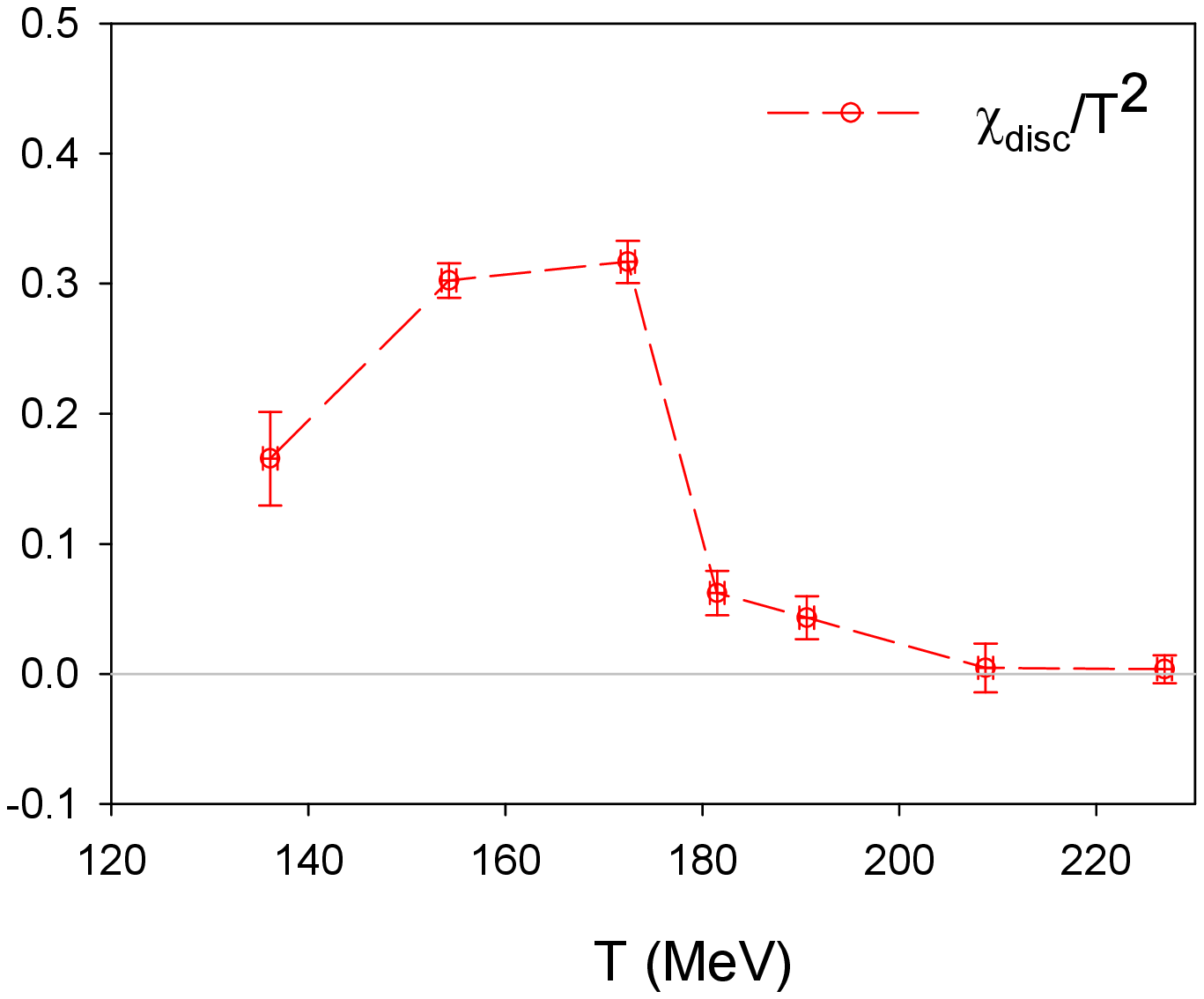}
&
\includegraphics*[height=4cm,width=4.8cm,clip=true]{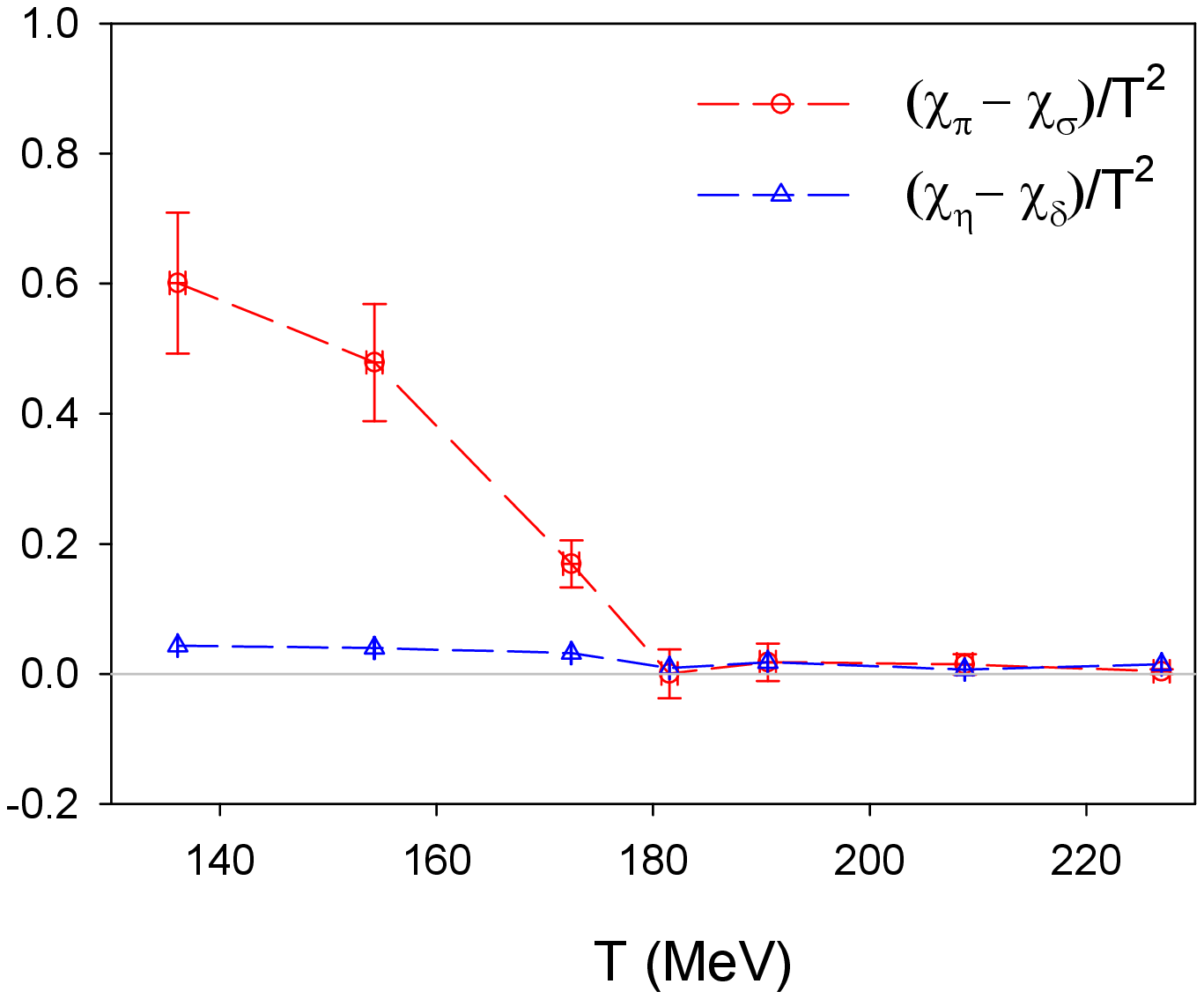}
&
\includegraphics*[height=4cm,width=4.8cm,clip=true]{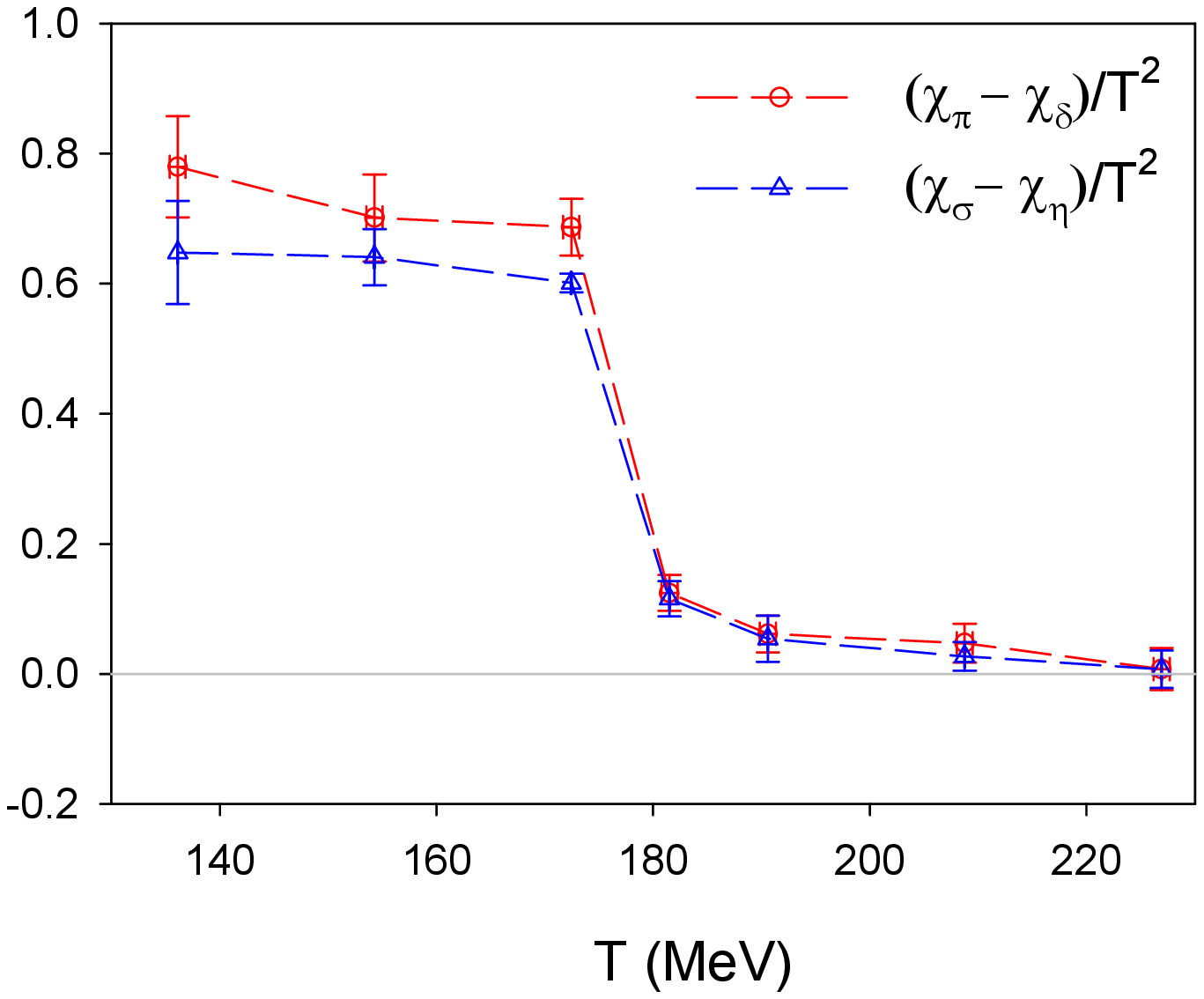}
\\
(a) & (b) & (c)
\end{tabular}
\caption{Differences of chiral susceptibilities versus $T$: (a) to locate the pseudo-critical temperature $ T_c $,  
(b) to probe the restoration of chiral symmetry, (c) to probe the restoration of $ U(1)_A $ symmetry.}
\label{fig:chis}
\end{center}
\end{figure}

For each configuration, we use 240 $ Z_2 $ noise vectors with dilution 
in the color and the Dirac spaces ($ 20 \times 3 \times 4 = 240 $) to compute the chiral susceptibilities for each configuration, 
and then obtain the average for each gauge ensemble.
The chiral susceptibility in the chiral limit at each temperature is obtained by linear extrapolation 
with 2-4 data points at $ m_{sea} a = 0.01, 0.015, 0.02, 0.03 $. 
In Fig. \ref{fig:chis} (a), we plot the dimensionless quantities 
$ \chi_{disc}/T^2 = (\chi_\sigma - \chi_{\delta})/(2 T^2) $ versus $ T $,  
and identify its peak as the pseudo-critical temperature, $ T_c \sim 172 $ MeV. 
In Fig. \ref{fig:chis} (b), we plot $ (\chi_\pi - \chi_\sigma )/T^2 $ and $ (\chi_\eta - \chi_\delta)/T^2 $ versus $ T $. 
It is evident that for $ T > T_c $, both of them go to zero, which implies that the chiral symmetry $ SU(2)_L \times SU(2)_R $
is restored for $ T > T_c $, even though the magnitude of $ (\chi_\eta - \chi_\delta)/T^2 $ is much smaller than 
$ (\chi_\pi - \chi_\sigma )/T^2 $. 
In Fig. \ref{fig:chis} (c), we plot $ (\chi_\pi - \chi_\delta )/T^2 $ and $ (\chi_\sigma - \chi_\eta)/T^2 $ versus $ T $.
We also observe that both of them decrease rapidly as $ T $ is slightly higher than $ T_c $, which implies that 
$ U(1)_A $ is restored at $ T_1 \simeq T_c $.       

Since the difference of any two chiral susceptibilities can be written in terms of the spectral density of $ D_c $, e.g.,  
\bea
\chi_\pi - \chi_\delta = \int_{0}^{\infty} d \lambda_c \frac{ 4 m_q^2 \rho(\lambda_c)}{(m_q^2 + \lambda_c^2)^2},  
\eea 
which is dominated by the low-lying eigenmodes of $ D_c $,    
we can examine the restoration of the chiral symmetry and the axial $ U(1) $ symmetry with 
the spectral density $ \rho(\lambda_c) $ of the low-lying modes.        
To this end, we calculate the zero modes plus (180+180) 
conjugate pairs of the lowest-lying eigenmodes of the overlap Dirac operator,  
for each configuration in the gauge ensembles. 
Our procedures have been outlined in Ref. \cite{Chiu:2011dz}.

\begin{figure}[!htb]
\begin{center}
\begin{tabular}{@{}cccc@{}}
\includegraphics*[height=4cm,width=5.3cm,clip=true]{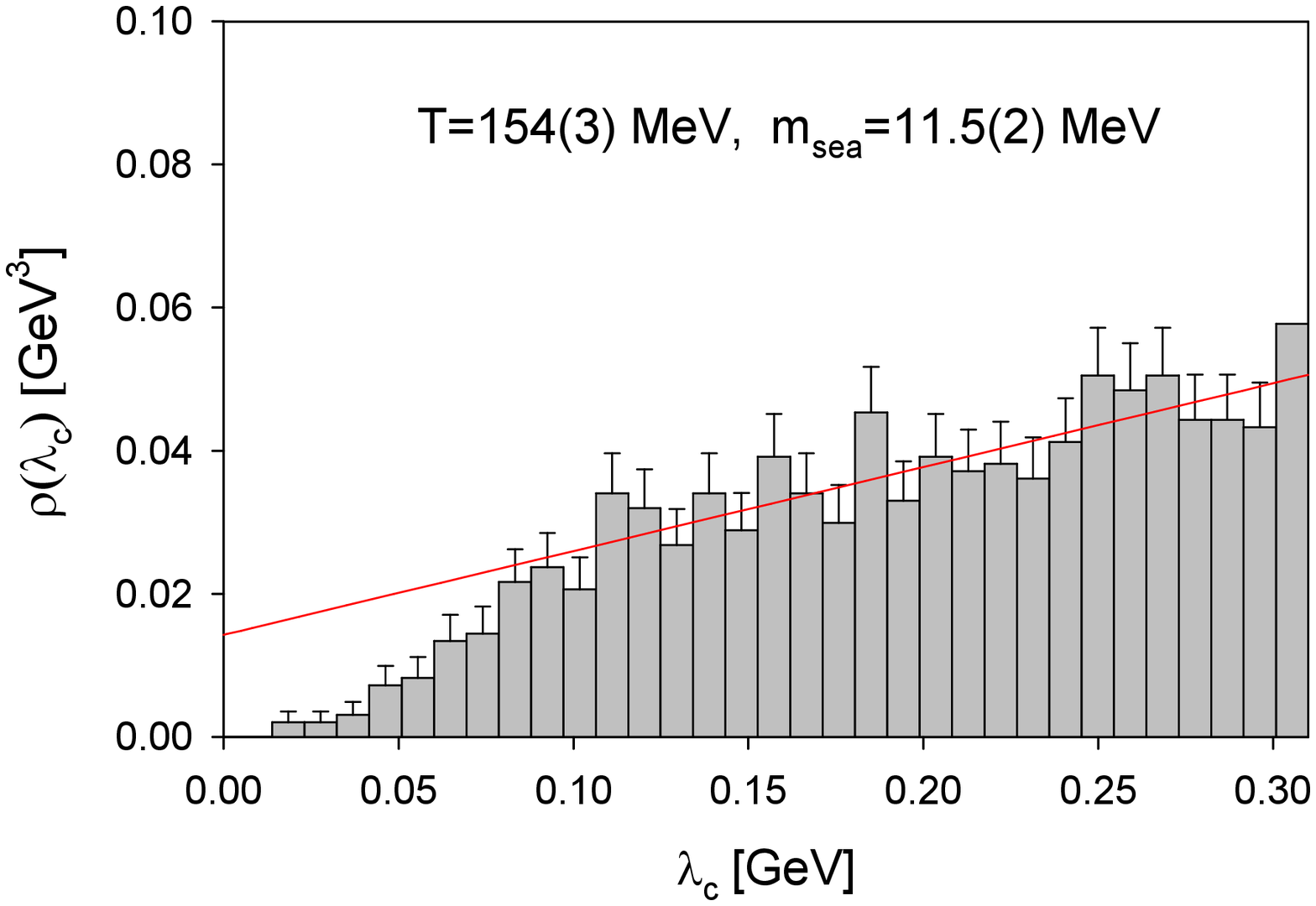}
&
\includegraphics*[height=4cm,width=4.8cm,clip=true]{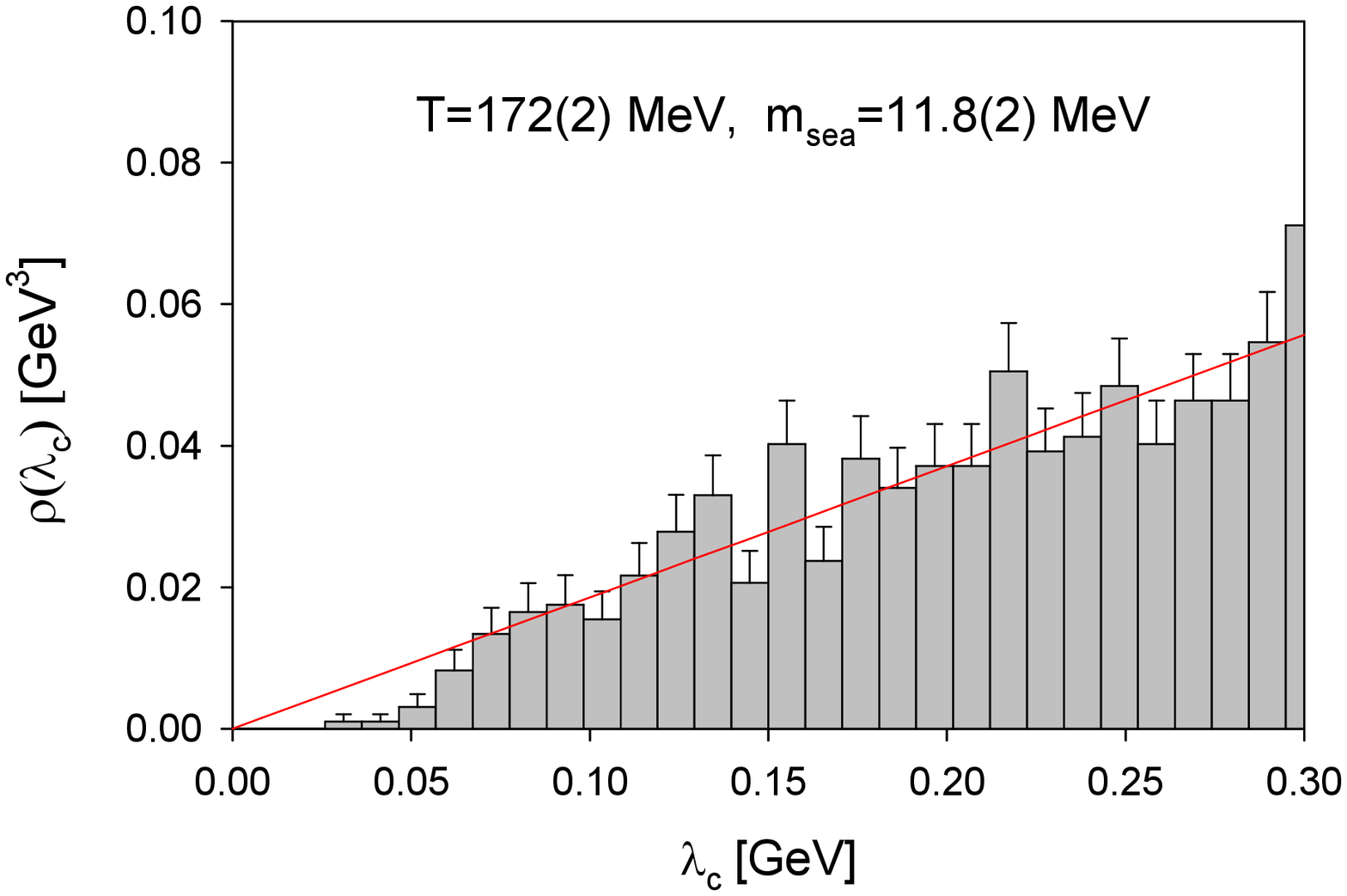}
&
\includegraphics*[height=4cm,width=4.8cm,clip=true]{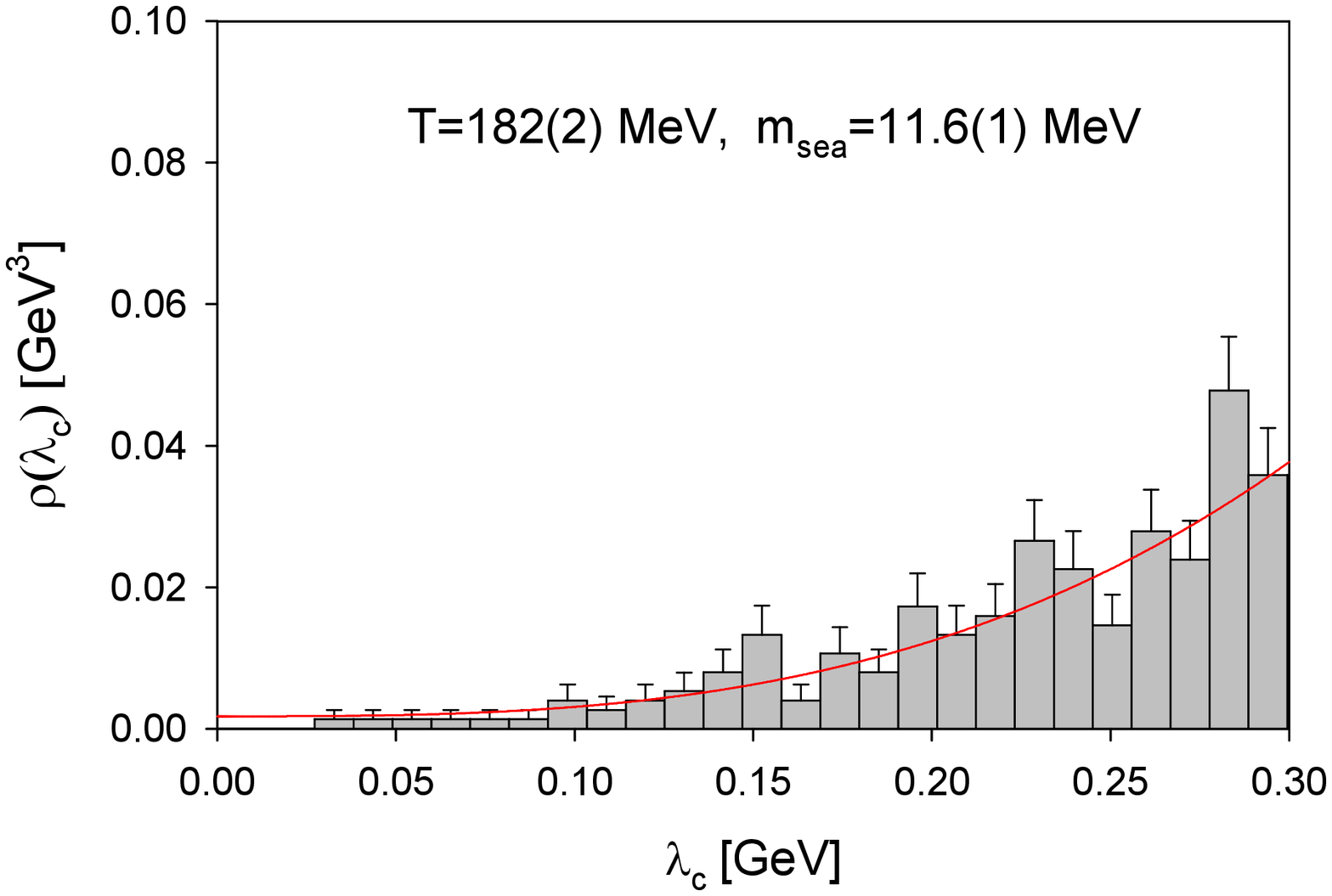}
\\
(a) & (b) & (c)
\end{tabular}
\caption{The distribution of the eigenvalue of $ (-iD_c) $: (a) $ T < T_c $, (b) $ T \simeq T_c $, (c) $ T > T_c $.
In (a) and (b), the red line denotes the linear fit for $ \lambda_c \in [ 0.1, 0.3 ] $ GeV, 
while in (c) the cubic fit $ c_0 + c_3 \lambda_c^3 $ for $ \lambda_c $ in the same range.}
\label{fig:rho_c}
\end{center}
\end{figure}

In Fig \ref{fig:rho_c}, we plot the eigenvalue density $ \rho(\lambda_c) $ versus $ \lambda_c $, 
for $ T $ [MeV] = 154(3), 172(2), and 182(2) respectively, with corresponding sea 
quark masses $ m_{sea} $ [MeV] = 11.5(2), 11.8(2), and 11.6(1), where the residual masses have been  
taken into account. 
At $ T = 153(3) $ MeV, the eigenvalue distribution in the interval $ [0.1, 0.3] $ GeV is 
well fitted by a linear function with nonzero intercept. The nonzero intercept implies that 
the chiral condensate is non-vanishing and the chiral symmetry is broken at this temperature. 
At $ T = 172(2) $ MeV, the eigenvalue distribution in the interval $ [0.1, 0.3] $ GeV is 
well fitted by a linear function with zero intercept. This suggests that 
this temperature is close to the pseudo-critical temperature $ T_c $. 
At $ T = 182(2) $ MeV, for $ \lambda_c $ in the interval $ [0.1, 0.3] $ GeV,  
the eigenvalue distribution cannot be fitted by linear or quadratic
function, but well fitted by the cubic function $ c_0 + c_3 \lambda_c^3 $ with $ c_0 \simeq 0 $.
This suggests that the chiral symmetry and the $ U(1)_A $ symmetry are both restored at this temperature, 
consistent with the degeneracies of the chiral susceptibilities ($ \chi_\pi \simeq \chi_\delta \simeq \chi_\eta \simeq \chi_\sigma $) 
as shown in Fig. \ref{fig:chis} (b) and (c). We also note that the suppression
of the low modes at $ T=182(2) $ MeV > $ T_c $ satisfies $ \rho(\lambda_c) = c_3 \lambda_c^3 $, consistent 
with the theoretical constraint obtained in Ref. \cite{Aoki:2012yj} for the restoration of 
the $ U(1)_A $ symmetry in the chirally symmetric phase.

\section{Concluding remarks}

Our preliminary results of the chiral susceptibilities and 
the eigenvalue density of the 4D effective Dirac operator 
in two flavors lattice QCD with optimal domain-wall fermion
suggest that the chiral symmetry and the $ U(1)_A $ symmetry 
are likely to be restored at nearby temperatures, $ T_1 \gtrsim T_c $, in the chiral limit. 
This implies that the chiral phase transition in two flavors QCD in 
the chiral limit could be first order,  
or second order in the $ U(2) \times U(2) / U(1)_V $ universality class \cite{Pisarski:1983ms}. 
A more precise determination of $ T_c $ and $ T_1 $, with a finer scan in $ T $, 
and also with larger volumes, are necessary to clarify this issue. 

%
  This work is supported in part by the National Science Council
  (Nos.~NSC99-2112-M-002-012-MY3,~NSC102-2112-M-002-019-MY3,~NSC99-2112-M-001-014-MY3,~NSC102-2112-M-001-011) 
  and NTU-CQSE (No.~102R891404). We also thank NCHC and NTU-CC for providing facilities to perform part of our calculations. 


\begin{thebibliography}{99}

\bibitem{Kaplan:1992bt}
D.~B.~Kaplan,
Phys.\ Lett.\ B {\bf 288}, 342 (1992);
Nucl.\ Phys.\ Proc.\ Suppl.\  {\bf 30}, 597 (1993).

\bibitem{Neuberger:1997fp}
  H.~Neuberger,
  Phys.\ Lett.\ B {\bf 417}, 141 (1998); 
%
%
  R.~Narayanan and H.~Neuberger,
  Nucl.\ Phys.\ B {\bf 443}, 305 (1995).


\bibitem{Bazavov:2012qja} 
  A.~Bazavov {\it et al.}  [HotQCD Collaboration],
  Phys.\ Rev.\ D {\bf 86}, 094503 (2012)

\bibitem{Cossu:2013uua} 
  G.~Cossu, S.~Aoki, H.~Fukaya, S.~Hashimoto, T.~Kaneko, H.~Matsufuru and J.~Noaki,
  Phys.\ Rev.\ D {\bf 87}, 114514 (2013)

\bibitem{Borsanyi:2012xf} 
  S.~Borsanyi, Y.~Delgado, S.~Durr, Z.~Fodor, S.~D.~Katz, S.~Krieg, T.~Lippert and D.~Nogradi {\it et al.},
  Phys.\ Lett.\ B {\bf 713}, 342 (2012)

\bibitem{Chiu:2002ir}
  T.~W.~Chiu,
  Phys.\ Rev.\ Lett.\  {\bf 90}, 071601 (2003);
%
  Phys.\ Lett.\ B {\bf 552}, 97 (2003);
%
  Phys.\ Lett.\ B {\bf 716}, 461 (2012);
%

\bibitem{Chiu:2009wh}
  T.~W.~Chiu {\it et al.}  [TWQCD Collaboration],
  PoS {\bf LAT2009}, 034 (2009); 
%
  T.~W.~Chiu [TWQCD Collaboration],
  J.\ Phys.\ Conf.\ Ser.\  {\bf 454}, 012044 (2013)

\bibitem{Chen:2012jya} 
  Y.~-C.~Chen and T.~W.~Chiu [TWQCD Collaboration],
  Phys.\ Rev.\ D {\bf 86}, 094508 (2012)

\bibitem{Chiu:2002eh}
  T.~W.~Chiu, T.~H.~Hsieh, C.~H.~Huang and T.~R.~Huang,
  Phys.\ Rev.\  D {\bf 66}, 114502 (2002).


\bibitem{Chiu:2011dz}
  T.~W.~Chiu, T.~H.~Hsieh and Y.~Y.~Mao [TWQCD Collaboration],
  Phys.\ Lett.\  B {\bf 702}, 131 (2011).


\bibitem{Chiu:2011bm}
  T.~W.~Chiu, T.~H.~Hsieh and Y.~Y.~Mao  [TWQCD Collaboration],
  Phys.\ Lett.\ B {\bf 717}, 420 (2012)


\bibitem{Luscher:2010iy} 
  M.~Luscher,
  JHEP {\bf 1008}, 071 (2010)

\bibitem{Borsanyi:2012zs} 
  S.~Borsanyi, S.~Durr, Z.~Fodor, C.~Hoelbling, S.~D.~Katz, S.~Krieg, T.~Kurth and L.~Lellouch {\it et al.},
  JHEP {\bf 1209}, 010 (2012)

\bibitem{Aoki:2012yj} 
  S.~Aoki, H.~Fukaya and Y.~Taniguchi,
  Phys.\ Rev.\ D {\bf 86}, 114512 (2012)


\bibitem{Pisarski:1983ms} 
  R.~D.~Pisarski and F.~Wilczek,
  Phys.\ Rev.\ D {\bf 29}, 338 (1984).

\end{thebibliography}
\end{document}